\documentclass[10pt,journal,compsoc]{IEEEtran}
\usepackage{comment}
\usepackage{enumitem} 

\usepackage{cite}
\ifCLASSINFOpdf
  \usepackage[pdftex]{graphicx}
\else
\fi
\begin{document}
%
\title{Move or Push? Studying Pseudo-Haptic Perceptions Obtained with Motion or Force Input}
%
%
%
%

\author{Yutaro~Hirao,~
        Takuji~Narumi,~
        Ferran~Argelaguet,~
        and~Anatole~Lécuyer 
\IEEEcompsocitemizethanks{
\IEEEcompsocthanksitem Y. Hirao is with Nara Institute of Science and Technology (NAIST), Nara 630-0192, Japan. E-mail: yutaro.hirao@is.naist.jp.
\IEEEcompsocthanksitem T. Narumi is with the University of Tokyo, Tokyo 113-8654, Japan. E-mail: narumi@cyber.t.u-tokyo.ac.jp.
\IEEEcompsocthanksitem A. Lécuyer and F. Argelaguet work with Univ. Rennes, Inria, IRISA, CNRS, Rennes, France.}
\thanks{}}


%
%

\markboth{Journal of}
{Hirao \MakeLowercase{\textit{et al.}}: Comparing Pseudo-Haptic Perceptions with Motion or Force Input}
%



\IEEEtitleabstractindextext{%
\begin{abstract}
Pseudo-haptics techniques are interesting alternatives for generating haptic perceptions, which entails the manipulation of haptic perception through the appropriate alteration of primarily visual feedback in response to body movements. However, the use of pseudo-haptics techniques with a motion-input system can sometimes be limited.
This paper investigates a novel approach for extending the potential of pseudo-haptics techniques in virtual reality (VR). The proposed approach utilizes a reaction force from force-input as a substitution of haptic cue for the pseudo-haptic perception. The paper introduced a manipulation method in which the vertical acceleration of the virtual hand is controlled by the extent of push-in of a force sensor. Such a force-input manipulation of a virtual body can not only present pseudo-haptics with less physical spaces and be used by more various users including physically handicapped people, but also can present the reaction force proportional to the user's input to the user. We hypothesized that such a haptic force cue would contribute to the pseudo-haptic perception. Therefore, the paper endeavors to investigate the force-input pseudo-haptic perception in a comparison with the motion-input pseudo-haptics. The paper compared force-input and motion-input manipulation in a point of achievable range and resolution of pseudo-haptic weight. The experimental results suggest that the force-input manipulation successfully extends the range of perceptible pseudo-weight by 80\% in comparison to the motion-input manipulation. On the other hand, it is revealed that the motion-input manipulation has 1 step larger number of distinguishable weight levels and is easier to operate than the force-input manipulation. 

\end{abstract}

}

\maketitle

\IEEEdisplaynontitleabstractindextext

%
\IEEEpeerreviewmaketitle

\IEEEraisesectionheading{\section{Introduction}\label{sec:introduction}}

%
%
%
%

\IEEEPARstart{I}{n} recent years, significant advancements have been made in virtual reality (VR) technology, allowing users to experience a sense of actual presence through visual and audio feedback. However, compared to audio-visual information, haptic feedback still possesses a narrower range of expression. To address this limitation, various approaches have been proposed to present haptic feedback. One intriguing method is the employment of pseudo-haptics techniques, which modify haptic perception by appropriately adjusting visual feedback in response to body movements \cite{Pusch2011, Ujitoko2021}. Examples of pseudo-haptics techniques include modifying the virtual force of a spring \cite{Lecuyer2000} or the perceived weight of an object \cite{taima2014controlling, rietzler2018breaking, Samad2019} by altering the control-display gain according to the user's motion within a virtual environment. The main advantage of the pseudo-haptics technique is that it primarily relies on visual stimuli to convey haptic perceptions, eliminating the need for bulky haptic devices. 

Traditionally, the pseudo-haptics technique for VR interaction involves the use of motion-input manipulation, where the spatial positions of the physical body are correlated with their virtual counterparts. However, There are several limitations with pseudo-haptics techniques with motion-input system. For example, a conventional pseudo-haptics technique reduces the virtual motion compared to the physical motion to present a stronger force perception and therefore, it requires larger physical motion and spaces. Moreover, because motion-input system in pseudo-haptics studies has morphologically equivalent input-output body mapping, it cannot be used by users with disabilities that prevent performing the required virtual motions. In addition to them, even when motion gains are applied and visual cues are altered, because there is no haptic feedback, the haptic cues remain unchanged from the absence of motion gains. This leads to a mismatch between the haptic estimates (e.g, width or weight of objects) derived from each sensory cue. Here, it is said that such separation between the estimates from each sensory cue can lead to discomfort or disrupt haptic perception when the disparities are too significant \cite{Pusch2011, rietzler2018breaking, honda2013imposed}. 

To address the above problems, Hirao et al. examined pseudo-haptics with a joystick-based avatar manipulation and confirmed that it can present pseudo-haptic weight perception with a similar level of the one of motion-input system \cite{hirao2020comparing}. Such a morphologically incongruent avatar manipulation system can present pseudo-haptics with less physical spaces and be used by more various users including physically handicapped people. Moreover, since the physical and virtual bodies are not consistent at the beginning, the problem of discomfort caused by the mismatch between them can less arise. 

\begin{figure}[tb]
 \centering
 \includegraphics[width=\columnwidth]{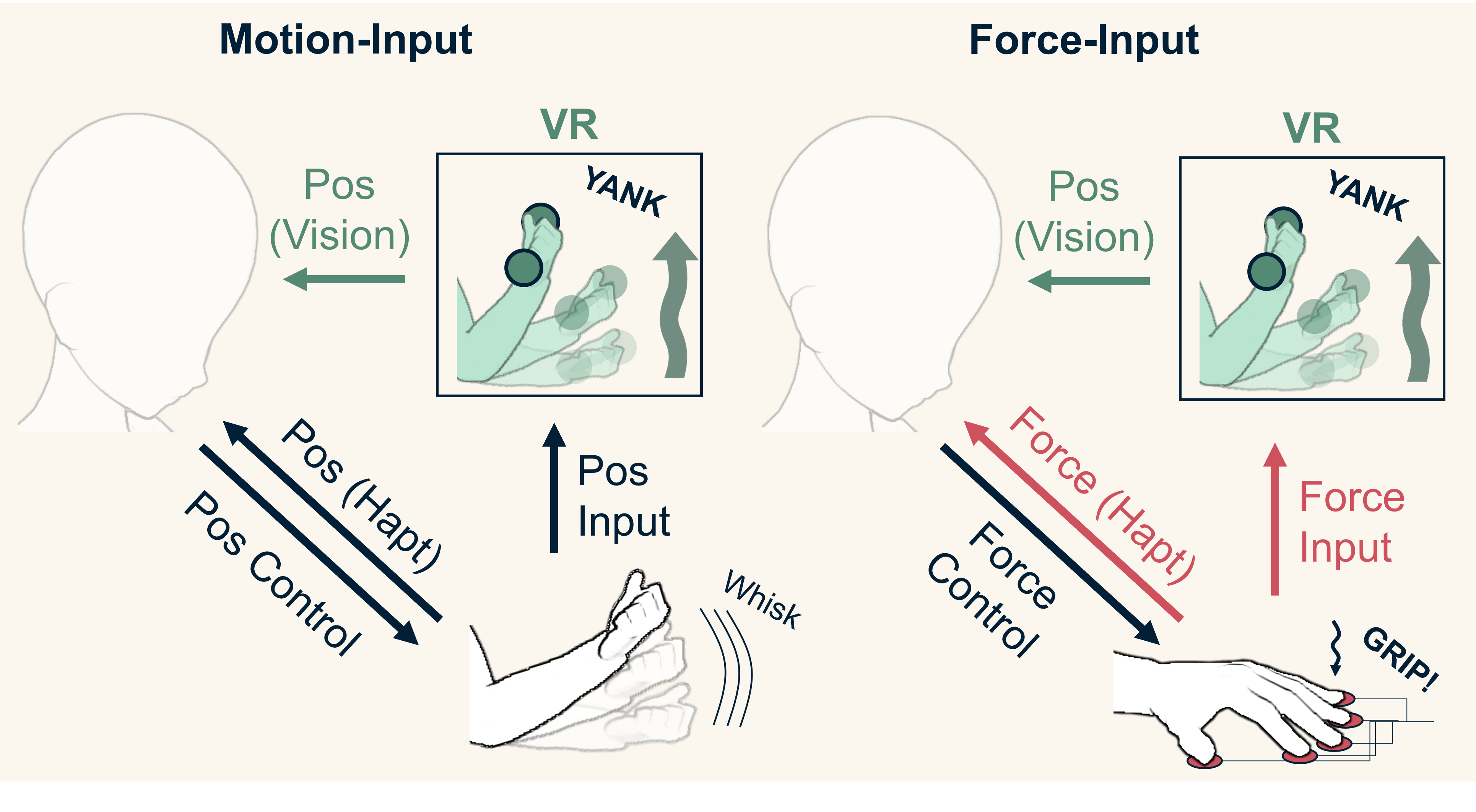}
 \caption{The concept diagram of the motion-input and force-input pseudo-haptics. The left indicates the motion-input and the right indicates the force-input manipulation. In the motion-input pseudo-haptics, users estimate haptics by combining visual posture/motion cue and haptic posture/motion cue. On the other hand, we hypothesize that in the force-input pseudo-haptics, users estimate haptics by combining visual posture/motion cue and the reaction force cue with the force-input.}
 \label{substitution}
\end{figure}

The purpose of this study is to further investigate a novel approach for extending the potential of pseudo-haptics. This paper investigates the pseudo-haptics with a force-input avatar manipulation that has morphologically incongruent input-output mapping. Such a force-input system not only has the merit above but also provides the user with a force feedback equivalent to the input force applied to the sensor when manipulating the virtual body. Hence, modifying the motion gain of the system not only changes the visual cues but also the haptic force cue to the user.
For instance, when lifting a virtual object, pseudo-weight perception can be produced by reducing the amount of visual movement relative to the actual input. In this scenario, the force-input manipulation method demands greater force to lift a heavier virtual object by the same extent as a lighter object.
Then, the paper endeavors to investigate if such a haptic force cue can be used as a substituted cue, i.e., be combined with a visual cues and contribute to the haptic estimates as a haptic force cue, even though the body parts used for the input and output are different. 
Fig. \ref{substitution} depicts a conceptual diagram of the difference of motion-input and force-input pseudo-haptics.
In this paper, we have decided to refer to this type of pseudo-haptics as FISpH (Force-Input Substituted Pseudo-Haptics), pronounced the same as "fish."


The paper reports two experiments which compare the pseudo-haptics utilizing the avatar manipulation method of force-input with that of motion-input. The paper investigates the pseudo-weight perception while virtual lifting task. Then, we developed a manipulation system in which the vertical acceleration of the virtual hand was controlled by the extent of push-in of a force sensor where the reaction force proportional to the user's input was fed back to the user (Fig. \ref{substitution}). The proposed system was compared to the motion-input manipulation system with respect to the range and resolution of presentable pseudo-weight perception. The first experiment assessed the number of levels of pseudo-haptic weights that could be presented by determining the range of gains that could be applied in each method and the resolution of the pseudo-haptic presentation within that range. The second experiment investigated the gain at which the pseudo-weight perception with the proposed manipulation was perceived to be equivalent to that produced at the maximum/minimum gain of the motion-input manipulation, in order to compare the two methods on the same axis of pseudo-weight perception.

The remainder of this paper is structured as follows. First, Section \ref{sec_relatedwork} introduces related work. It provides the overview of the pseudo-haptics techniques and reviews the studies on haptic substitution. Then, Sections \ref{experiment} describes the two experiments to compare the effectiveness of FISpH and motion-input pseudo-haptics. Subsequently, Section \ref{discussion} provides a comprehensive discussion of all experimental results. Specifically, it discusses the effect of the two manipulation methods on pseudo-weight perception and usability. Moreover, it also introduces the discussion of the experimental methods for measuring pseudo-haptics. Subsequently, it mentions the limitations of the paper and future work.
Finally, Section \ref{sec_conclusion} concludes the paper. We would like to note that all experiments in the paper were approved by the local ethical committee.

\section{Related Work}
\label{sec_relatedwork}

\subsection{Pseudo-Haptics Techniques}
\label{LR:pseudo-haptics}
Since the first paper on pseudo-haptics in 2000 \cite{Lecuyer2000}, the pseudo-haptics techniques have been studied for various haptic perceptions. Haptic properties of objects can be classified into roughly two categories \cite{jones2006human} and many pseudo-haptics techniques are investigated to modify these properties: material properties, such as texture (such as roughness and friction \cite{okamoto2012psychophysical}; e.g. \cite{narumi2017resistive, ujitoko2019presenting, fukushima2013study, lecuyer2004feeling}), compliance (e.g. \cite{Lecuyer2000, punpongsanon2015softar, Argelaguet2013, kumar2017mechanics, tatezono2009effect}), and weight (e.g. \cite{taima2014controlling, rietzler2018breaking, Samad2019}); and geometric properties, such as curvature (e.g. \cite{ban2012modifying}), orientation and angle (e.g. \cite{kohli2010redirected, azmandian2016haptic, ban2012modifying, han2018evaluating}), and size (e.g. \cite{ban2013modifying}). Moreover, the retargeting and redirection research \cite{kohli2010redirected, razzaque2005redirected} can be also considered as one part of pseudo-haptics research because the geometric properties are explained as haptic properties and it is mentioned that the modification of the haptic properties using multisensory (mainly visual and haptics) integration processing is a pseudo-haptics technique (\cite{Pusch2011}). 

However, if we look at the quantitative results of to what extent these techniques can modify haptic perception, the pseudo-haptics techniques still have a large gap for the application. To elaborate, psychophysical experiments suggest that the pseudo-haptics techniques can alter weight perception roughly by 3~5\% \cite{taima2014controlling, Samad2019}, compliance perception by 20~35\% \cite{lecuyer2001boundary, tatezono2009effect}, friction perception by 23\% \cite{ujitoko2019presenting}, and fine roughness perception by 5\% \cite{ujitoko2019modulating}. Moreover, studies including retargeting and redirection field suggest that the pseudo-haptics techniques can alter hand motion perception up to 13.75\% in the forward, and up to 6.18\% in the backward direction without being noticed \cite{zenner2019estimating}. In this way, several studies suggest that the pseudo-haptics techniques can alter haptic perception roughly by 5\% to 35\% depending on the target haptic perception. However, the range of perceptual intensity that can be modified is still narrow compared to using haptic devices, and it cannot be said enough when considering various practical use in VR applications. 

One of the possible reasons for this limitation is related to the fact that the pseudo-haptics technique separates visual and haptic information. Here, larger gaps are necessary to induce more intensive perception; however, gaps that are too large result in discomfort or a failure to integrate the information \cite{Pusch2011, rietzler2018breaking, honda2013imposed}. Based on Pusch and L{\'e}cuyer's model \cite{Pusch2011}, larger discrepancies could result in larger complements and substitutions of sensory information from personal memories and experiences, which would increase individual differences. Moreover, if the gap is too large, the information is determined to be coming from a different source and is not integrated \cite{ernst2004merging}. Furthermore, forward and inverse dynamics theory can provide another explanation in that the illusion would not arise when the brain no longer has a model to which the large gap can be attributed \cite{honda2013imposed}. 

Although this limitation of pseudo-haptics is well-known, there are few studies on extending the applicable range of the pseudo-haptics techniques.
First, addition of multiple cues is suggested to enhance the pseudo-haptics although they did not mention to what extent it can improve \cite{hirao2018augmented, punpongsanon2015softar}. Then, Ban and Ujitoko proposed a method for solving the discrepancy between visual and physical position by connecting the discrepancy between a user's finger and the cursor on a touchscreen with a visual string \cite{ban2018enhancing}. 
Another possible approach involves increasing the reliability of visual information in sensory integration, i.e., to decrease the variance of visual information in MLE so that the final perception is more dependent on visual information by wider field of view in VR \cite{williams2019estimation} or more realistic avatar \cite{ogawa2020effect}. 
However, these approaches do not manipulate the haptic information, thus the pseudo-haptics is still largely restricted by the haptic cue from the body.
On the other hand, some studies endeavor to add noise signals to haptic cues so that the final perception is less dependent on haptic cues by using noisy galvanic vestibular stimulation (GVS) \cite{matsumoto2021redirected}, bone-conducted vibration (BCV) \cite{weech2017vection, kondo2023effects}, or tendon vibration \cite{hirao2023leveraging}. For example, Hirao et al. confirmed that the tendon vibration as noise on haptic motion cues can extend the range of visual/physical discrepancy without being noticed by about 13\%. However, these approaches has unintended side effects that might have effect on the user experiences or does not have large improvement.
Compared with these approaches, the proposed method, FISpH, endeavors to tackle the problem from completely novel angle, which introduces new body schema having force-input manipulation and newly designs the integration law of visual and haptic cues.

\subsubsection{Pseudo-haptics using force-input system}
There are several studies investigating pseudo-haptics with force-input devices. Actually, the very first research on pseudo-haptics uses force-input interface that mimics an interaction with a virtual spring (that looks like a syringe) on a screen \cite{Lecuyer2000}. Most of these studies endeavor to simulate virtual compliance or stiffness by changing the gain of force-input to virtual motion with a touch display \cite{ridzuan2012direct}, a rod type device \cite{heo2019pseudobend}, or handheld devices that sensor the grabbing or pinching force \cite{matsumoto2017displaying, yabe2017pseudo, lee2019torc, adilkhanov2020vibero}. These approaches can be said straightforward and strong because such compliance or stiffness has the push-in force parameter and motion parameter in their physical law and the pseudo-haptics techniques of changing their gain are rational. 
Moreover, these system use the same body schema in physical and virtual reality, i.e., the body part that is used to the force-input and the one that moves in VR is the same. For example, TORC simulates virtual hand's pinch movement by a real hand's pinch force-input \cite{lee2019torc}.
On the other hand, the force-input system in this paper uses the different body schema in physical and virtual reality, i.e., the height of a virtual hand is manipulated by the extent of push-in of a force sensor with an index finger. Moreover, this paper investigates weight perception and its physical law does not explicitly have the push-in force parameter in its physical equation. Therefore, unlike the previous studies, it is not obvious that the force feedback, that is, the reaction force while the push-in, functions as a haptic cue for the weight estimates, combined with the visual motion cue. In other words, this paper endeavors to introduce a new body schema in virtual body manipulation and substitute the reaction force cue of the finger for the somatosensory cue of the arm to strengthen the effect of the pseudo-haptics techniques.

\subsection{Sensory Substitution}
As the FISpH can be said to substitute original haptic cue (haptic cues expected from virtual body part) with a new one (haptic cues from force-input), this section reviews the research of haptic substitution.
The study of how to substitute haptic feedback has been examined mainly in the field of the research on sensory feedback of prostheses \cite{antfolk2013sensory, svensson2017review}. The non-invasive methods to present sensory feedback of prostheses are technically devided into mechanotactile feedback (e.g. \cite{antfolk2012sensory, antfolk2013transfer, panarese2009humans, kim2009design, sensinger2009examination, marasco2011robotic}), vibrotactile feedback (e.g. \cite{pons2005objectives, mondal2009upgrading}), electrotactile feedback (e.g. \cite{xu2015effects, mulvey2009use, franceschi2016system}), auditory feedback (e.g. \cite{lundborg1999hearing, gonzalez2012psycho}), and hybrid of them (e.g. \cite{clemente2014novel}) \cite{svensson2017review}.

Another way to classify the feedback methods is from the perspective of if the sensory feedback is modality-matched to (i.e. is felt in the same modality as) the target sensory input \cite{antfolk2013sensory, svensson2017review}. For example, studies on the sensory feedback of upper limb prostheses often try to substitute pressure exerted on prosthetic fingers during grasping movements. In this case, the mechanotactile feedback is modality-matched feedback, and the other sensory substitutions such as vibrotactile, electrotactile, and auditory feedback are not. In general, modality-matched feedback is more intuitive and requires less cognitive effort during manipulation \cite{antfolk2013sensory} than non-modality-matched feedback, and can improve the sense of embodiment to the prostheses \cite{marasco2011robotic, svensson2017review, ehrsson2008upper} as well as enables closed-loop control \cite{antfolk2012sensory, panarese2009humans}.

Then, which body part should the substituted feedback be presented to? In the field of prostheses research, it is often presented at the location directly related to the prosthesis manipulation (e.g. body-powered prostheses \cite{antfolk2013sensory}), the part which has similar physiological characteristics (e.g. \cite{panarese2009humans}), or the part where the amputee felt the "phantom sensation of the lost body part" or which has undergone "targeted reinnervation surgery." It is known that brain plasticity often induces a sensory remapping of the lost body part on the different body parts such as the amputation stump \cite{bjorkman2012phantom, hunter2005dissociation} and face \cite{flor1995phantom, farne2002face}, where the amputee can feel a sense of touch of the lost body part. Moreover, targeted reinnervation is a technique that "allows severed cutaneous nerves to reinnervate skin on a different portion of the body" \cite{sensinger2009examination}. Therefore, it is natural and rational to present sensory substituted feedback to these body parts because it is physiologically correct \cite{antfolk2013sensory}.
Meanwhile, in recent years, sensory substitution techniques have been investigated also for non-amputee users due to the increasing demand for haptic feedback in VR and augmented reality research. For example, Pezent et al. proposed a wrist band shaped device named Tasbi which can present squeeze and vibrotactile feedback to the wrist \cite{pezent2019tasbi}. They proposed to substitute vibrotactile feedback for fingertip contact sensation with virtual objects or user interfaces, and squeeze feedback for weight or stiffness associated with manipulating virtual objects. Moreover, Kameoka et al. proposed a system named Haptopus, that transforms the touch sensation of the hand into tactile (suction) feedback to the face \cite{kameoka2018haptopus}. Furthermore, Hiki et al. proposed a sensory substitution system that transforms the pressure sensation of the hand to the one of the sole of the foot \cite{hiki2018substitution}. Their experimental results suggest that their system can allow participants to recognize the weight, stiffness, and shape of an object in a similar capability to the one using a hand.
These techniques are valuable because they can present haptic information to users while allowing them hands-free interaction.
However, unlike in most cases of the prostheses studies for amputees, these techniques do not present substituted feedback in a physiologically correct way. Moreover, these approaches do not strictly substitute/remove the original haptic cue but use another haptic cue in addition to the original one. As a result, the original haptic estimates from the body part, such as the fingers, remain present. Thus, the haptic estimates is still restricted by the one derived from the original haptic cue. For example, when haptic feedback of the hand is substituted with other body parts to present the perception of the hardness of a virtual object interacting with a virtual hand, even though the information of the hardness can be interpreted, the perception of the hardness and/or user experience may be interfered by the fact that the hand is physically grasping empty. This may be solved by long-term learning with brain plasticity \cite{kieliba2021robotic}, but that is not obvious at this moment.
Then, this paper proposes to develop a new body schema that can reduce/remove the influence of haptic estimates derived from the original haptic cues, and that substitutes modality-matched feedback for the original one.

\section{Experiments}
\label{experiment}
This section compares FISpH, which substitutes the original haptic cues with the reaction force cues induced by force-input manipulation in a new body schema, with the conventional motion-input pseudo-haptics.
The main focus of the experiment is to investigate the pseudo-haptics of the two methods from a view of the applicable range and the resolution of the pseudo-haptic perception. Specifically, the experiments investigate pseudo-weight perception while lifting a virtual object.

The research questions in the experiments follow:

\begin{itemize}[leftmargin=1cm]
    \item [\textrm{[RQ1]}] Which manipulation method has wider range of pseudo-weight perception?
    \item [\textrm{[RQ2]}] Which manipulation method has more number of levels of pseudo-weight perception?
    \item [\textrm{[RQ3]}] Is the current implementation of force-input manipulation controllable?
\end{itemize}

\subsection{XP1: How many levels of pseudo-haptic weight can motion-input and force-input manipulation present?}
The first experiment investigated the range of applicable motion gain out of which users start to feel the interaction is incoherent/weird or lose the feeling of control. Moreover, this also allows to investigate the resolution of pseudo-haptic weight perception with each manipulation. With these results, i.e., the range and resolution, we calculated how many levels of pseudo-haptic weight each manipulation can present.

\subsubsection{Apparatus}

\begin{figure}[ht]
 \includegraphics[width=\columnwidth]{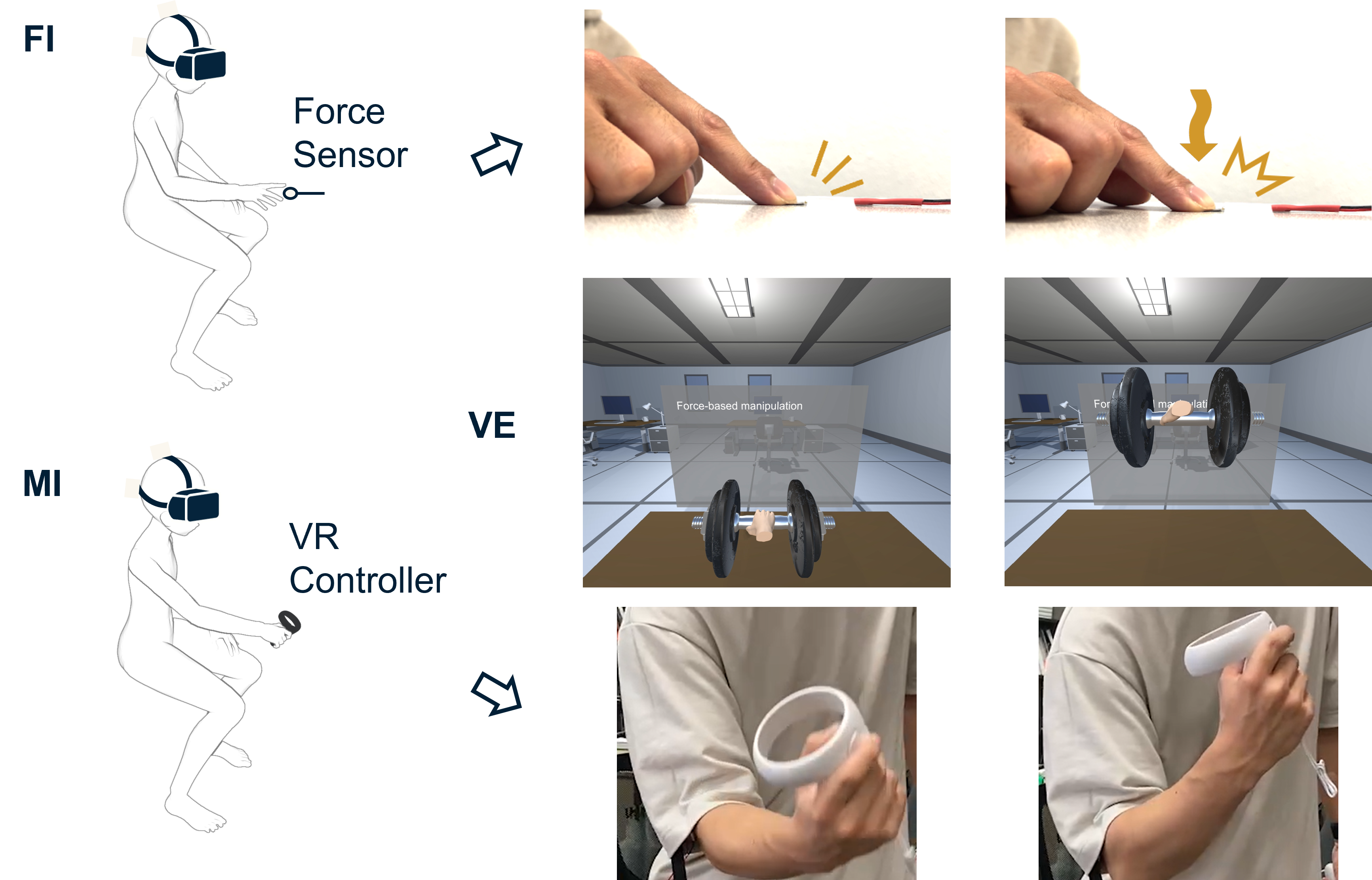}
 \centering
 \caption{The apparatus and virtual environment. The top figures indicate the force-input manipulation and the bottom figures indicate the motion-input manipulation. The middle figures indicate the virtual environment.}
 \label{sub_study2_apparatus}
\end{figure}

Figure \ref{sub_study2_apparatus}  shows the experimental apparatus and virtual environment. The experiment was conducted with participants in a sitting position. Participants wore a VR headset (Oculus Quest 2). A force sensor (FSR 400, Interlink Electronics Inc.) was attached to a table beside the participants. In VR, there was a low table and a virtual dumbbell on it in front of the participants. The virtual dumbbell was placed at 0.7 m in height from the ground. 

In this experiment, there were two manipulation methods for a virtual hand. One manipulation method was the motion-input manipulation. Here, participants held a VR controller with their right hand. The position of the virtual hand tracked that of a VR controller, and the posture of the virtual hand was set to correspond to the physical hand when participants grabbed the VR controller. Participants could grab and release the virtual dumbbell by using an index trigger. Once they grabbed the virtual dumbbell, the virtual hand and dumbbell moved only on a vertical axis and their height tracks VR controller's height with a gain.

The other manipulation method was the force-input manipulation. Here, participants put their right arm on the table beside them and press the force sensor with their index finger. In this method, we calculated the acceleration of the virtual hand and object by applying force on the sensor and Newton’s motion equation. First, we obtained an equation for the relationship between output voltage from the force sensor [V] ($V_{out}$) and applied force [N] ($F_s$) on it (equation \ref{eq:appliedForce}) from a data sheet. Here, 0.20 in equation \ref{eq:appliedForce} indicates the area of the contact surface of the force sensor. Then, for the method of simulating virtual force ($F_v$) induced by the virtual hand from the applied force ($F_{sensor}$), this time, we determined to simply multiply a constant of 10 by applied force to the force sensor and a gain ($G$) that is used for pseudo-haptics (equation \ref{eq:virtualforce}). Note that there would be a space to design a better simulation of virtual force from the applied force. However, this study is at the first step to investigating the effect of force-input manipulation on pseudo-haptics and does not focus on the proper design of force-input manipulation. Therefore, we determined to use the most simple equation by using a constant of 10 that is decided by our pilot test. Subsequently, the acceleration of virtual hand/object [$m/s^2$] ($A_v$) was calculated by Newton's motion equation (equation \ref{eq:dynamics}). Here, the weight of the virtual object was set to 0.15 [kg] which is the standard weight of VR controller. In the force-input manipulation, the virtual hand was set to the grab position of the virtual dumbbell, and they moved in a vertical axis when participants apply force to the sensor.

\begin{equation}
F_s = (0.067 \cdot e^{1.5 \cdot V_{out}}) \cdot 0.20 \cdot G ~~~~ [N]
\label{eq:appliedForce}
\end{equation}

\begin{equation}
F_v = F_s \cdot 10 \cdot G ~~~~ [N]
\label{eq:virtualforce}
\end{equation}

\begin{equation}
A_v = \frac{F_v}{0.15} - 9.8 ~~~~ [m/s^2]
\label{eq:dynamics}
\end{equation}

\subsubsection{Task}
There were three phases in the experiment. The first phase was the "learning task phase" which aimed to let participants get used to each manipulation and investigate the task performance. Here, participants conducted reaching tasks on the vertical axis with each manipulation method. A target height was presented with a blue bar. Participants grabbed and lifted up a virtual object to the blue bar. After the participants reached the target bar and released the object, then, the bar, virtual hand, and virtual object disappeared. Subsequently, the virtual hand and object appeared in an original position, and the next target bar was presented. The target bar was randomly positioned in a range from 0.2 m to 0.7 m range from the table height (0.7 m) in front of the participants. This reaching task took 15 seconds and the participants conducted 2 repetitions of this 15 seconds trial for each manipulation method. The participants were asked to complete the task as fast and precisely as possible (tried not to go over the blue bar but to stop and release at the bar’s height).

The second phase was the "range task phase" which aimed to investigate the applicable gain for pseudo-haptics in each manipulation. The staircase method was used for this aim. Here, the applicable gain for pseudo-haptics indicates a minimum and a maximum gain where participants start to feel discomfort or lose the feeling of control. As reported in the related work section, the most common range of "applicable gain" in previous work in the field of redirection or pseudo-haptics is the one under the detection threshold of visual/physical discrepancy. This range is the most strict range where users do not notice the visual modification by a gain. However, we consider that range of applicable gain in practice is wider than this range, where users can notice the visual/physical discrepancy but feel pseudo-haptic perception without feeling unnaturalness for the interaction. we consider that this wider definition of the range of applicable gain is more practical. Indeed, there are many studies on pseudo-haptics that uses a gain beyond the detection threshold of visual/physical discrepancy (e.g. \cite{taima2014controlling, rietzler2018breaking, Samad2019}). Moreover, because we investigate pseudo-haptics with force-input manipulation and the force-input manipulation does not have an obvious definition of the detection threshold of visual/physical discrepancy, we cannot use the range using the detection thresholds. Therefore, in this experiment, we determined the range of applicable gain as the one where users do not feel discomfort or do not lose the feeling of control. 

In the task of the staircase method, the range of adjustable gain was from 1.0 to 3.0 for the high gain group (gain $>$ 1) and from 0.3 to 1 for the low gain group (gain $<$ 1). Each gain group had ten equal steps i.e., a step value in the high gain group was 0.2 and that in the low gain group was 0.07. The task was a free interaction with the virtual dumbbell in 5 seconds in VR. After the free interaction, participants answered the question of “It was too heavy/light.” by "Agree" or "Disagree" by touching a virtual button in VR. Here, an experimenter explained that the question indicates that “I felt it was too heavy/light and start to feel weird or lose the feeling of control." A staircase method had two series of trials whose initial gain was either the minimum or maximum gain of the range of adjustable gain. When participants answered “Agree”, the gain moved one step closer to a gain of 1.0 and vice versa. When each series in a staircase method reached 5 turning points, the experiment was over.

The third phase was the "resolution task phase" which aimed to investigate the resolution of pseudo-haptic perception. The method of constant stimuli was used for this aim. Participants compared the sense of the weight of reference and comparison dumbbell within one of the manipulation and answer which dumbbell was heaver. The reference gain was 1.0 and the comparison gains were 0.76, 0.84, 0.92, 1.08, 1.16, 1.24 (-24\%, -16\%, - 8\%, +8\%, +16\%, +24\%, respectively). The task was to interact with two virtual dumbbells for 3 seconds for each and answered a question: “Which dumbbell did you feel heavier?” by “The first” or “The second” by touching a button in VR. Each comparison gain was compared to the reference gain ten times. Therefore, participants conducted 120 trials in total ($6 \: comparison \: gains \times 2 \: manipulations \times 10 \: repetitions$). 

\subsubsection{Collected data}
In the learning task, the overshoot distance from the target height, each reaching duration, and the number of reaching the target were recorded. This data was used to check if the force-input manipulation was usable for this lifting task. 
In the range task, the minimum and maximum applicable gain was obtained by averaging adjusted gains at 10 turning points in the staircase method.
In the resolution task, JND of pseudo-haptic weight perception was calculated by psychometric function obtained by the method of constant stimuli.

\subsubsection{Procedure}
Twenty participants (fifteen males and five females in their twenties) participated in the experiment. First, the objectives, methods, procedures, and how each manipulation works were explained to them. Next, the participants completed a consent form. Then, the participants were asked to wear the VR headset. Here, the participants also checked the position of the force sensor on a table. Before the main tasks, participants practiced each manipulation method with a virtual dumbbell with a gain of 1. Subsequently, the participants conduct 3 main tasks. The combination of the order of the manipulation method in each task was counterbalanced between the participants. Furthermore, other orders such as that of the high/low gain group in the second phase and comparison gains in the third phase were randomized. The participants could take a break whenever they wanted. After the experiment, the participants were remunerated with an Amazon gift card of 5 euros for their participation. The total duration of the whole procedure was around 45 minutes.

\subsubsection{Results}

\begin{table}[b]
\caption{The results of task scores in the first phase. The scores at the second trial are used. The values indicate $average\pm Standard Error$.}
\label{task score}
\begin{center}
\begin{tabular}{|c||c|c|}\hline 
Manipulation & \begin{tabular}{c} Number of Reaching \end{tabular} & \begin{tabular}{c} Overshoot (m) \end{tabular}  \\ \hline \hline
motion-input & $9.4 \pm 0.5$ & $0.06 \pm 0.01$ \\ \hline
force-input & $9.0 \pm 0.6$ & $0.23 \pm 0.03$ \\ \hline
\end{tabular}
\end{center}
\vspace*{-3mm}
\end{table}

\begin{table}[b]
\caption{The results of the applicable gain threshold.
Note that the results of motion-input and force-input are not directly comparable because the meaning of gain is different. The values indicate $average\pm Standard Error$}
\label{threshold}
\begin{center}
\begin{tabular}{|c||c|c|}\hline 
Manipulation & \begin{tabular}{c} Threshold (Low) \end{tabular} & \begin{tabular}{c} Threshold (High) \end{tabular} \\ \hline \hline
motion-input & $0.57 \pm 0.04$ & $1.51 \pm 0.08$ \\ \hline
force-input & $0.46 \pm 0.03$ & $1.88 \pm 0.13$ \\ \hline
\end{tabular}
\end{center}
\vspace*{-3mm}
\end{table}

Table \ref{task score} shows the results of the task scores of the reaching task, i.e., number of reaching the bar within the trial and overshoot distance from the target height when the participants released the virtual object. 
Here, we used the values at the second trial. In addition, the results of the overshoot of each reaching task were averaged for each trial. Paired t-test was conducted for each number of reaching and overshoot. As a result, a significant difference was found in the results of overshoot ($p < 0.001, d=1.76$).

Table \ref{threshold} shows the results of the second phase. With these results, we can calculate the range of the applicable gains in each manipulation. The range of applicable gain for motion-input manipulation was 0.94 and that for force-input manipulation was 1.42.

\begin{figure}[t]
 \includegraphics[width=\columnwidth]{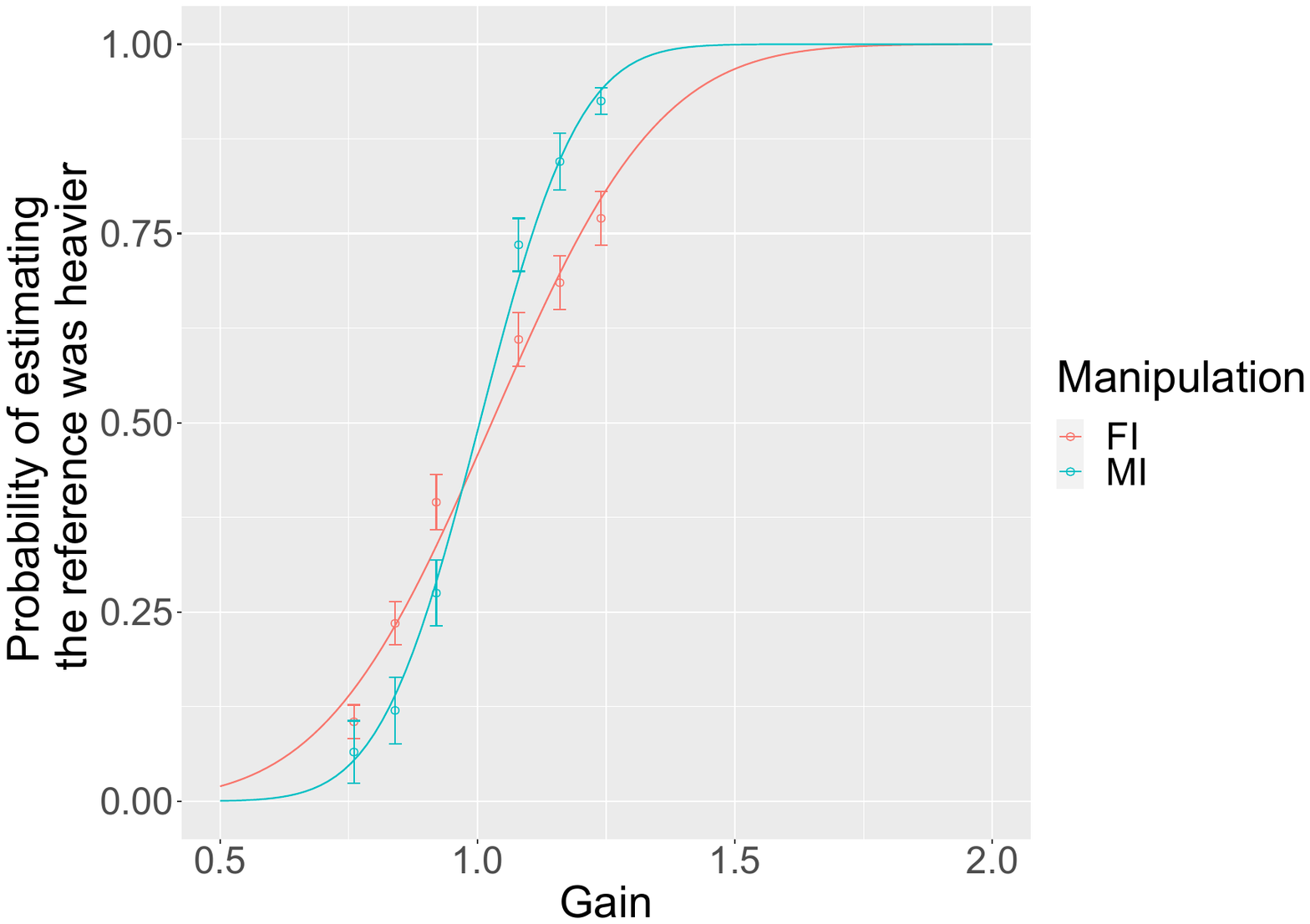}
 \centering
 \caption{The results of the third phase: Psychometric curves of the averaged results. It plots the average percentage of answers in which participants considered the reference dumbbell heavier than the comparison dumbbell. The error bar indicates the standard error.}
 \label{sub2_JND}
\end{figure}

Then, for the results of the third phase, the Probit analysis \cite{finney1971probit} was conducted for the averaged values of all participants' results, which calculated the parameters of the best-fitting cumulative normal function. Subsequently, we computed the JND for each manipulation as half of the distance between the points of 25\% and 75\% on the psychometric curve. These values were 0.10 for the motion-input manipulation and 0.17 for the force-input manipulation. Figure \ref{sub2_JND} shows the psychometric curves calculated with the averaged values of all participants' results. With the results of the second and third phases, we calculated the number of pseudo-haptic weight levels that each manipulation method can present. Then, it was 9 levels for motion-input manipulation and 8 for force-input. 

\subsection{XP2: What is the equality between the pseudo-haptic weight with the force-input and motion-input manipulation?}
\label{sec:sub2_xp2}
The second experiment investigated the point of subjective equality (PSE) of the pseudo-haptic weight perception between each manipulation. In the first experiment, we obtained the range of the applicable gain for pseudo-haptic weight and its JND with each manipulation. However, the gains cannot be directly compared because the meaning of the gains is different in each manipulation. Therefore, here, we investigated the PSE to enable the comparison of the results of the first experiment between each manipulation.

\subsubsection{Apparatus}


The apparatus was the same as that in the first experiment. However, this time, participants used both hands during the task: one hand was used for the motion-input and the other hand for the force-input manipulation.
The hand for each manipulation was switched during the experiment.

\subsubsection{Task}
\begin{figure}[ht]
 \includegraphics[width=\columnwidth]{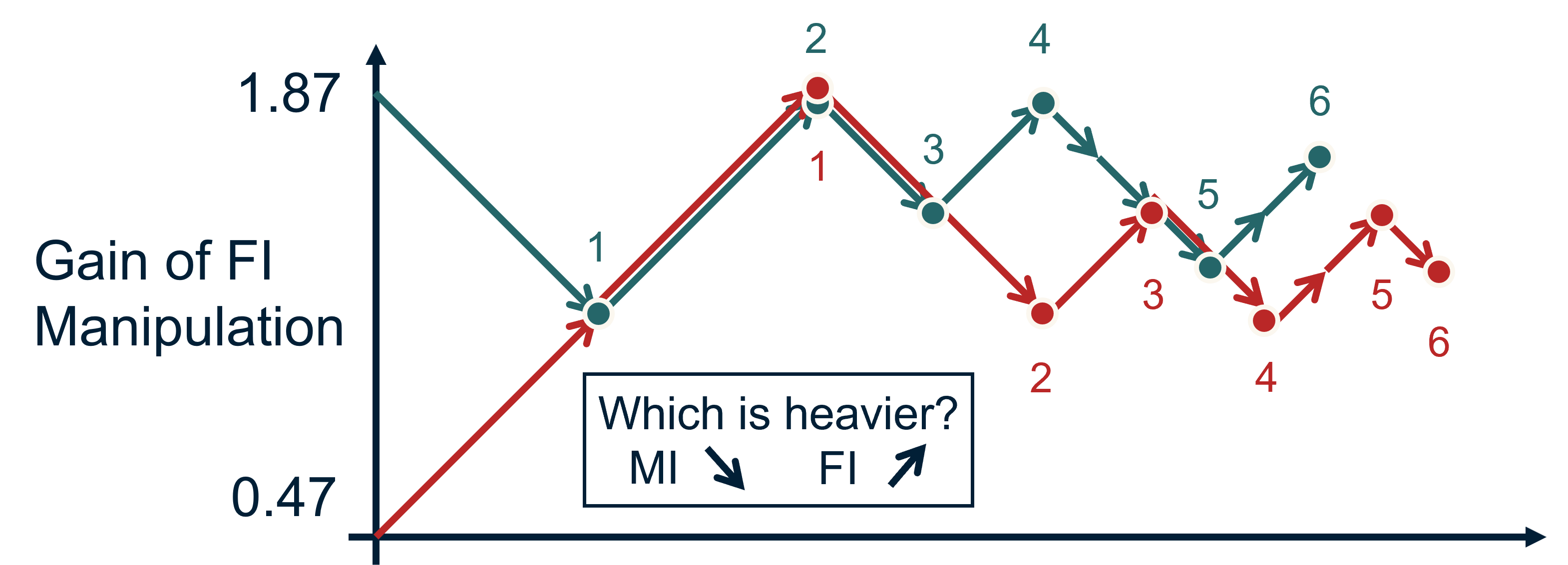}
 \centering
 \caption{The conceptual diagram of an adaptive staircase method. The arrow indicates one trial to the next trial and its length shows the distance of each step gain. The dots and numbers indicate the turning points where the answer of participants changed from the last one or the gain is likely to go beyond the maximum or minimum gain. The green arrows indicate descending series and the red indicates ascending series. The PSE was calculated as the average of gains at each turning point.}
 \label{sub2_xp2_staircase}
\end{figure}

An adaptive staircase method was used to investigate the PSE of the sense of weight between each manipulation. Specifically, the PSE for weight perception at a maximum and minimum gain in the motion-input manipulation was obtained by manipulating a gain in force-input manipulation. Participants used one hand for the motion-input manipulation and the other hand for the force-input manipulation to lift up a virtual dumbbell in VR. Then, the participants answered the question “which dumbbell did you feel heavier?” by “motion-input manipulation” or “force-input manipulation.” 
The reference dumbbell was with the motion-input manipulation at minimum or maximum gain obtained in the first experiment, i.e., 0.57 or 1.51, respectively. We determined the motion-input manipulation as the reference condition by referring to the observation of the first experiment where most of the participants experienced that the gain around the maximum/minimum threshold with the force-input manipulation was heavier/lighter than that with the motion-input manipulation.
The comparison dumbbell was with the force-input manipulation at a gain in the range of minimum and maximum gain obtained in the first experiment, i.e., 0.47 and 1.87, respectively.

Figure \ref{sub2_xp2_staircase} shows the conceptual diagram of the adaptive staircase method. The initial gain of the reference dumbbell was either maximum (0.47) or minimum (1.87). There were three levels of distance for each step for the adaptive staircase method. The minimum distance was 0.175 and the others were 0.35 and 0.7. These were 1/2/4 times the minimum distance. If the participants chose the force-input manipulation as a heavier dumbbell, the force-input gain moved one step closer to the maximum, and vice versa. Here, if the gain was likely to move beyond the range, it remained at the same value. The steps first changed with the maximum distance and the distance changed after two turning points. There were also ascending and descending series and each series ended at the sixth turning point. The participants repeated the staircase method two times for each maximum and minimum gain of the motion-input manipulation. Note that the combination of the participant's hands and the manipulations changed between the two repetitions.
At the end of the experiment, participants were asked which method was better in a point of sense of weight by a question of “With which manipulation method did you feel a better sense of weight/effort?”

\subsubsection{Collected data}
Gains at each turning point were recorded. Then, these gains were averaged to calculate PSE for each participant. Furthermore, the answer to the last question "With which manipulation method did you feel a better sense of weight/effort?" was saved.

\subsubsection{Procedure}
Eighteen participants (16 males and 2 females in their twenties excluding two in their 40s) participated in the experiment. The flow before the main task was completely the same as in the first experiment. After the practice trial, participants conducted the main task of the staircase method. When participants changed the combination of the hand and manipulation, they rotated their chair and VR scene was re-centered. The combination of the order of the maximum/minimum gain of the motion-input manipulation and the order of the hand-manipulation combination was counterbalanced between the participants. The participants could take a break whenever they wanted. After the experiment, the participants were remunerated with an Amazon gift card of 1000 yen for their participation. The total duration of the whole procedure was around 30 minutes.

\subsubsection{Results}

\begin{table}[h]
\caption{The PSE of pseudo-weight perception between the motion-input and force-input manipulation.}
\label{PSE}
\begin{center}
\begin{tabular}{|c||c|c|}\hline 
Manipulation &  ($Average \pm Standard Error$) \\ \hline \hline
motion-input Max (1.51)   & $1.57 \pm 0.07$       \\ \hline
motion-input Min (0.57)   & $0.78  \pm 0.06$       \\ \hline
\end{tabular}
\end{center}
\vspace*{-3mm}
\end{table}

Gains at each turning point were averaged to calculate the PSE for each participant. Then, paired t-test was conducted for the results between the combinations of hand and manipulation. As a result, no significant difference was found. Therefore, we determined to compute PSE by averaging the results of 2 repetitions (different hand-manipulation combinations). Table \ref{PSE} shows the results of the PSE.

As a result of the last question, 7 out of 18 participants mentioned that the motion-input manipulation provides a better sense of weight/effort, while the other 11 participants mentioned the opposite.


\section{Discussion}
\label{discussion}

\subsection{The force-input can have wider pseudo-weight range while the motion-input can have higher resolution}

\begin{figure}[h]
 \includegraphics[width=\columnwidth]{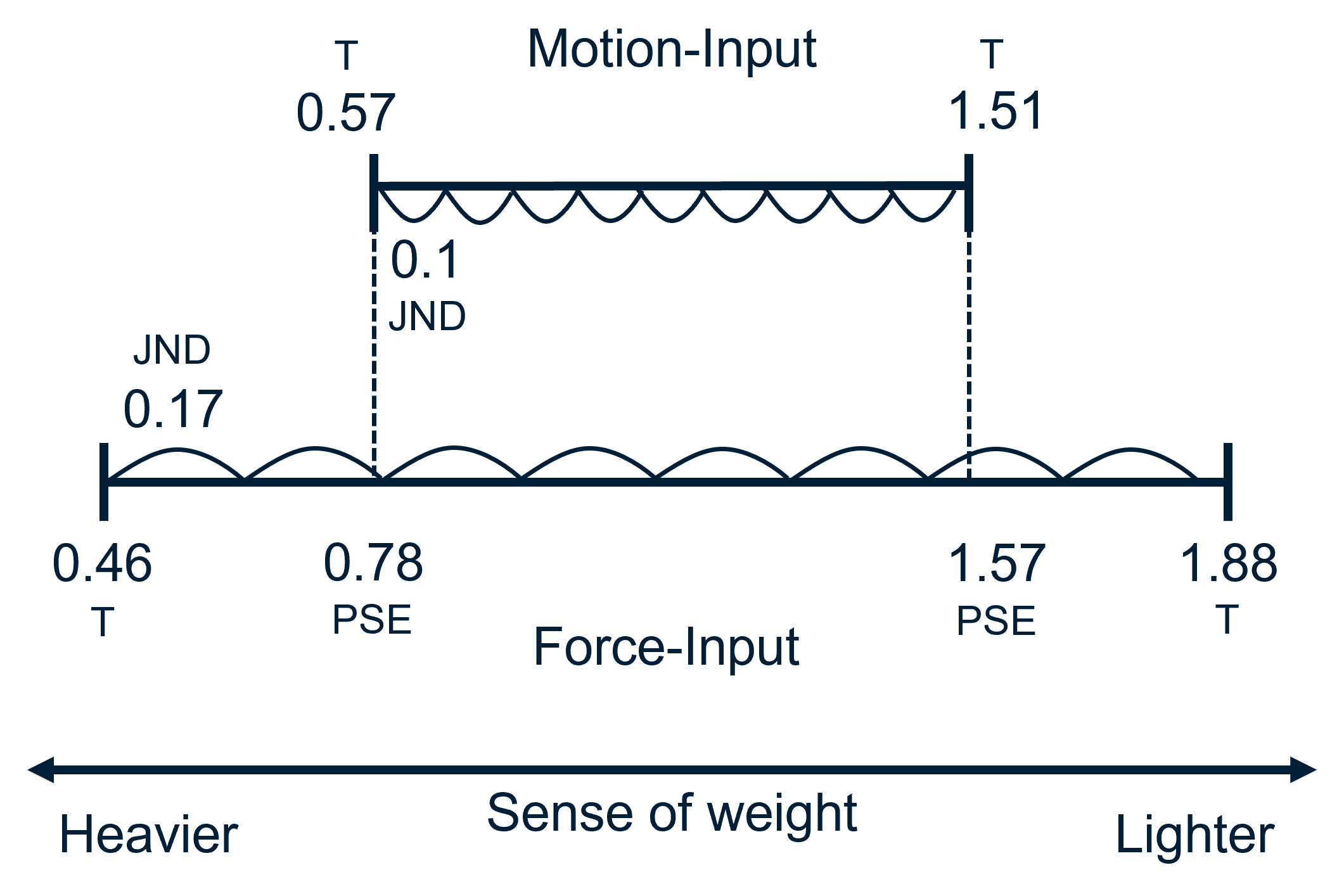}
 \centering
 \caption{The summary of the results of the two experiments. "T" indicates threshold. Here, the horizontal bars indicate the range of gain which can be used for presenting pseudo-weight perception with each manipulation method. The bars are scaled to make them comparable in the axis of the same pseudo-weight perception.}
 \label{sub2_generalResults}
\end{figure}

Figure \ref{sub2_generalResults} indicates the summary of the results of the two experiments. Taken together the results of the first and second experiments, the force-input manipulation can actually provide a wider range of pseudo-haptic weight perceptions (about 1.8 times wider), i.e. lighter and heavier perceptions. This results suggest that FISpH can have wider range of pseudo-weight perception (RQ1) and support the effectiveness of the proposed concept. It is noteworthy that these findings imply that users can experience varying degrees of pseudo-weight perception, even when interacting with significantly different body schema, such as controlling a virtual hand using pressure input from an index finger, without much training effort. This would suggest that people could integrate the substituted haptic cue with visual motion cue to estimate virtual weight.
One thing to note is that we cannot rule out the possibility that the results would be different if different formulas or implementation were used for the force-input manipulation. Nevertheless, this paper's main contribution is that it successfully suggested that the perceived intensity range of pseudo-haptics can be extended in both heavy and light directions with a given implementation, and confirmed the potential of this concept as the first step.

However, at a point of a number of levels of presentable pseudo-haptic weight, the motion-input manipulation can present 1 more level (9 levels) than the force-input (8 levels). Moreover, the difference in weight perception between each step is smaller in the motion-input manipulation than in the force-input. The JND of pseudo-weight perception was gain of 0.1 for motion-input and gain of 0.17 for force-input manipulation. These results conclude that motion-input can have more multiple levels of pseudo-weight perception (RQ2). 
Still, again, these results could be different by improving the implementation of force-input body manipulation.

\subsection{The motion-input is easier to operate}
Regarding usability, the force-input manipulation had a lower control compared to the motion-input manipulation as participants overshoot about 23 cm from the point they wanted to stop although the total number of reaching the target was not largely different. These findings imply that the current implementation of the force-input manipulation could be manipulable, but it may exhibit not small (approximately a 23 cm) margin of error when utilized for tasks requiring haste (RQ3).
However, this paper did not aim to design a control law from force input to output motion but focused on the investigation of the pseudo-haptic perception of the force-input manipulation as a first study. Therefore, future work should consider the improvement of the implementation of the force-input manipulation in a point of usability and sense of embodiment.
In summary, with the current implementation, it can be concluded that the motion-input manipulation should be employed when a small modification of weight perception is enough but better operability of manipulation is required, while the force-input should be employed when a larger range of weight perception is desired.

\subsection{The difficulties of measuring pseudo-haptics}
Although the paper tried to evaluate pseudo-haptic perception from a view of the threshold (range), JND (resolution), and PSE (equality) with psychophysical methods, how to evaluate pseudo-haptics is one of the biggest topics to be discussed and there is no consensus on this point yet. Then, this subsection reviews and discusses the experimental methods to evaluate pseudo-haptics, and concludes that our experimental method could be one of the promising solutions especially when comparing pseudo-haptics in very different methods.

First, the Likert scale asking participants to answer their pseudo-haptic perception from 1 to 7 can be done with a relatively simple experiment design and the experiment requires relatively little time. However, it cannot be clear to what extent the proposed method can extend the pseudo-haptics quantitatively. Therefore, it would be a good method as a first step of the research to see if the proposed method is effective or in cases where there are many experimental conditions, tasks, or questions.

Then, the intensity of pseudo-haptics can be evaluated by directly answering the intensity on a scale from 0 to 100 using VAS (e.g., \cite{hirao2020comparing}).
To directly answer the haptic sensation (VAS) may reflect the person's pure perception. However, the results tended to have a large variance because the evaluation axis in a mind may vary from person to person. Moreover, the participant's understanding of the question may cause their answers to be biased. In most cases, the design of proper wording in the question would be difficult. For example, when participants evaluate pseudo-haptics in a question about the sense of weight, the interpretation of "weight" can be different for each participant. Participants may evaluate the weight from their fatigue, sense of body motion, or other similar perception such as stickiness. Therefore, researchers should pay attention to the design of the question and explanation for the experiment.

Next, the behavioral indicators such as exerted force (gripping force) could capture the effect of pseudo-haptics (\cite{hirao2018augmented}). Although the results of the exerted force seem to reflect the intensity of pseudo-haptics at a certain level, the variances are relatively large and does not completely fit the subjective results. Therefore, further study is needed to determine how accurately the difference in pseudo-haptic perception can be captured in the behavioral factor or how to improve it.

Moreover, a psychophysical method where pseudo-haptics were evaluated by referring to physical quantities can also be used (e.g., \cite{taima2014controlling, Samad2019, hirao2020comparing}) At first glance, this method may seem to give the most robust results because this method does not need to suffer from wording issues. However, we need to keep in our mind that the pseudo-haptic perception may not be qualitatively equal to the corresponding haptic perception in our daily life \cite{lecuyer2009simulating, Ujitoko2021}. Moreover, the experimental results suggest that the results would be different depending on the experimental methods, that is, adjusting the visual gain against the physical quantities (e.g., \cite{hirao2020comparing}) or vice versa (e.g., \cite{taima2014controlling, Samad2019}). This can note that researchers need to pay attention to the experimental design not to underestimate the capability of pseudo-haptics techniques due to the experimental task.

Subsequently, in the paper, we tried to evaluate the pseudo-haptics from a view of the range with resolution and PSE. 
This information enables us to get the characteristics of the pseudo-haptics technique and compare the different techniques. However, it is difficult to determine what constitutes a range in pseudo-haptics. Moreover, it is also difficult to evaluate the thresholds because the evaluation method also affects the results \cite{grechkin2016revisiting}.
Here, there are two possible thresholds for the spatial limitation in pseudo-haptics: a detection threshold of visual/physical discrepancy (e.g., \cite{hirao2023leveraging}) and an acceptable threshold with noticing the discrepancy (e.g., \cite{rietzler2018breaking}). Therefore, it can be considered that there are four levels of pseudo-haptics experience: the first level is that users do not feel pseudo-haptics nor realize the pseudo-haptics technique (mostly the visual gain); the second level is that users feel pseudo-haptics but unaware of the technique; the third level is that users feel pseudo-haptics and aware of (accept) the technique; the fourth level is that users hardly feel pseudo-haptics or have large variances for the occurrence of pseudo-haptics and feel unnatural or discomfort for the technique.
In the conventional pseudo-haptic studies, the range was defined as one including the first and second levels because the psychophysical experiment can figure out relatively exact and robust threshold of the second and third levels. This method is probably the most strict and robust method for evaluating pseudo-haptics at present. However, because the third layer can be and is often used in actual applications \cite{rietzler2018breaking}, this strict evaluation would underestimate pseudo-haptics potential.
On the other hand, in the present paper, the range was defined as one including the first, second, and third levels. we consider that this method can have the most practical results that can be used in actual applications at present.
However, it is difficult to determine the threshold of the third and fourth levels. For example, the word "discomfort" has a wide range of meanings and can be interpreted ambiguously by each participant. Therefore, further study is required to investigate how to determine a proper threshold for the third and fourth levels.

\subsection{Limitations and Future work}
First, the paper focused on the pseudo-haptics and did not investigate other important factors of avatar manipulation such as sense of embodiment or more detailed task performance. Sense of embodiment plays an important role in sense of immersion/presence, i.e. sense of being there, in VR \cite{slater2010simulating}. In the future work, these factors should be considered and the implementation of the design of the force-input manipulation should be improved referring to the results. Specifically, the law of converting force into virtual body motion has a space to be improved. One thing to note is that the results obtained in the paper can be different with the different implementation. 
Moreover, another possible future work is to investigate the methodology to develop a force-input manipulation system of the whole virtual body. In the current experiment, the implementation of force-input manipulation was limited to the vertical motion of a virtual hand.

\section{Conclusion}
\label{sec_conclusion}
The paper investigated effectiveness of the force-input body manipulation on pseudo-weight perception in a comparison with that of the motion-input manipulation.
Specifically, the study investigated the force-input manipulation where the motion velocity of the virtual hand corresponds to the force applied to a force sensor in a comparison with motion-input manipulation. The experimental results suggest that the reaction force induced by the force-input could substitute the original haptic cue and be integrated with the visual motion cue, inducing pseudo-weight perception. Moreover, the force-input manipulation can extend the range of presentable pseudo-weight perception by 80\% compared to the motion-input manipulation, while the motion-input manipulation has 1 more steps of the presentable pseudo-weight levels. Taken together, with the current implementation, it can be concluded that the motion-input manipulation should be employed when a small modification of weight perception is enough but better operability of manipulation is required, while the force-input should be employed when a larger range of weight perception is desired.

\section*{Acknowledgments}
This work was partially supported by the MEXT Grant-in-Aid for Scientific Research (S) (19H05661), Grant-in-Aid for JSPS Fellows (21J12284), and Grant-in-Aid for Research Activity Start-up (23K20007).

\ifCLASSOPTIONcaptionsoff
  \newpage
\fi



\bibliographystyle{IEEEtran}
\bibliography{main}
%


%

\vspace*{-3\baselineskip}

\begin{IEEEbiography}[{\includegraphics[width=1in,height=1.25in,clip,keepaspectratio]{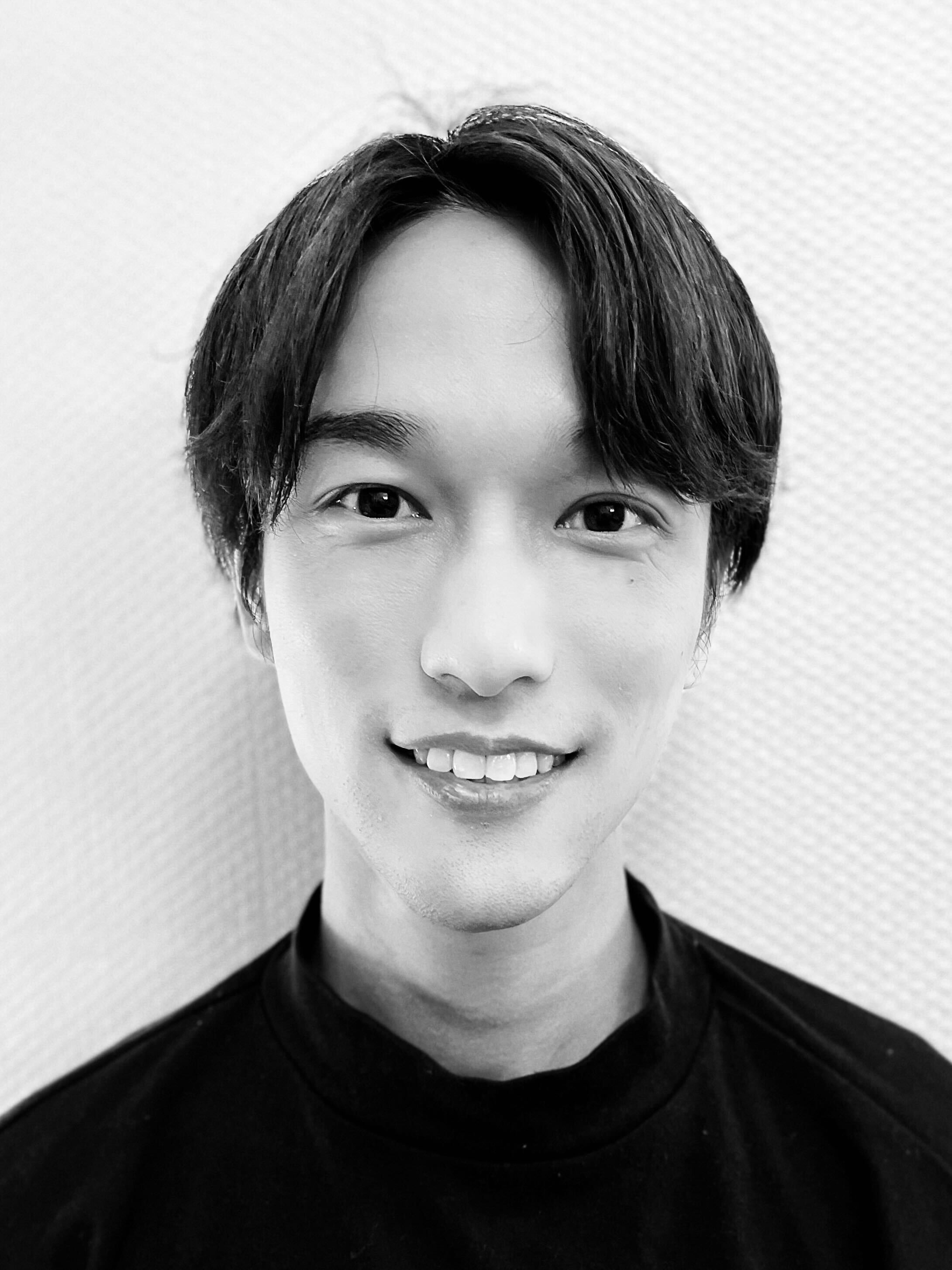}}]{Yutaro Hirao} is an Assistant Professor at Nara Institute of Science and Technology (NAIST), Japan (23-). His main research interests include virtual reality (VR), cross-modal interaction, haptic perception (pseudo-haptics), and embodiment. He received his B.S. and M.S. in engineering from Waseda University (18-20) in Japan, and his Ph.D. in information science and technology form the University of Tokyo (20-23).
\end{IEEEbiography}

\vspace*{-3.4\baselineskip}

\begin{IEEEbiography}[{\includegraphics[width=1in,height=1.25in,clip,keepaspectratio]{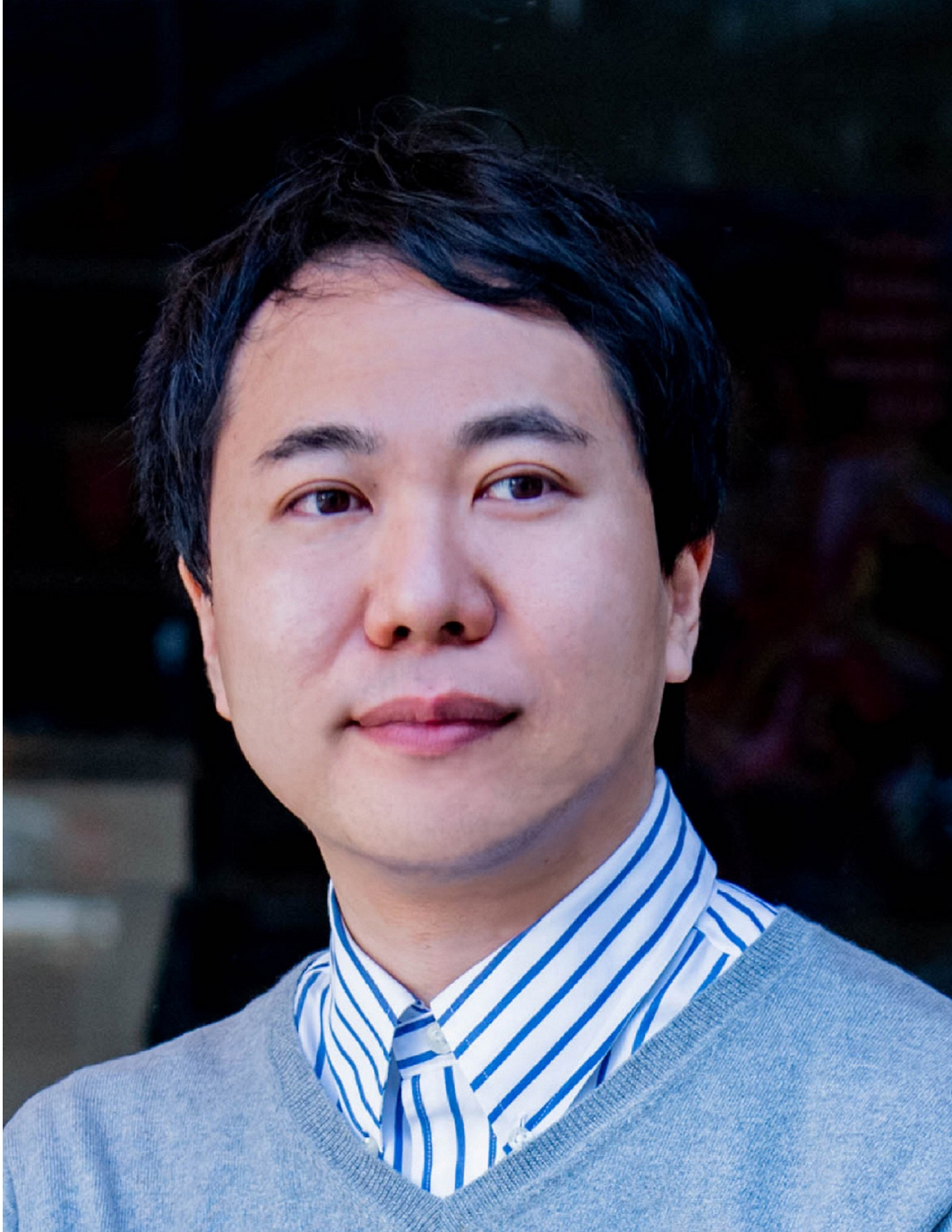}}]{Takuji Narumi}
is an associate professor at the Graduate School of Information Science and Technology, the University of Tokyo. His research interests broadly include perceptual modification and human augmentation with virtual reality and augmented reality technologies. He received BE and ME degree from the University of Tokyo in 2006 and 2008 respectively. He also received his Ph.D. in Engineering from the University of Tokyo in 2011.
\end{IEEEbiography}

\vspace*{-4\baselineskip}

\begin{IEEEbiography}[{\includegraphics[width=1in,height=1.25in,clip,keepaspectratio]{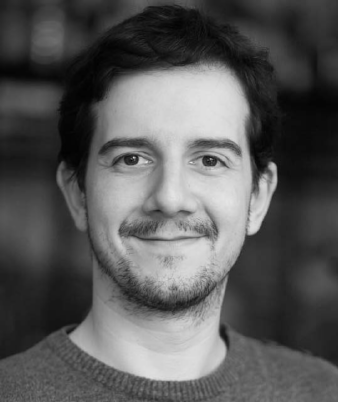}}]{Ferran Argelaguet}
is an Inria research scientist at the Hybrid team (Rennes, France) since 2016. He received his PhD degree from the Universitat Polit\`ecnica de Catalunya (UPC), in Barcelona, Spain in 2011. His main research interests include 3D user interfaces, virtual reality and human-computer interaction. He was program co-chair of the IEEE Virtual Reality and 3D User Interfaces conference track in 2019 and 2020, and the journal track in 2022.
\end{IEEEbiography}

\vspace*{-4\baselineskip}

\begin{IEEEbiography}[{\includegraphics[width=1in,height=1.25in,clip,keepaspectratio]{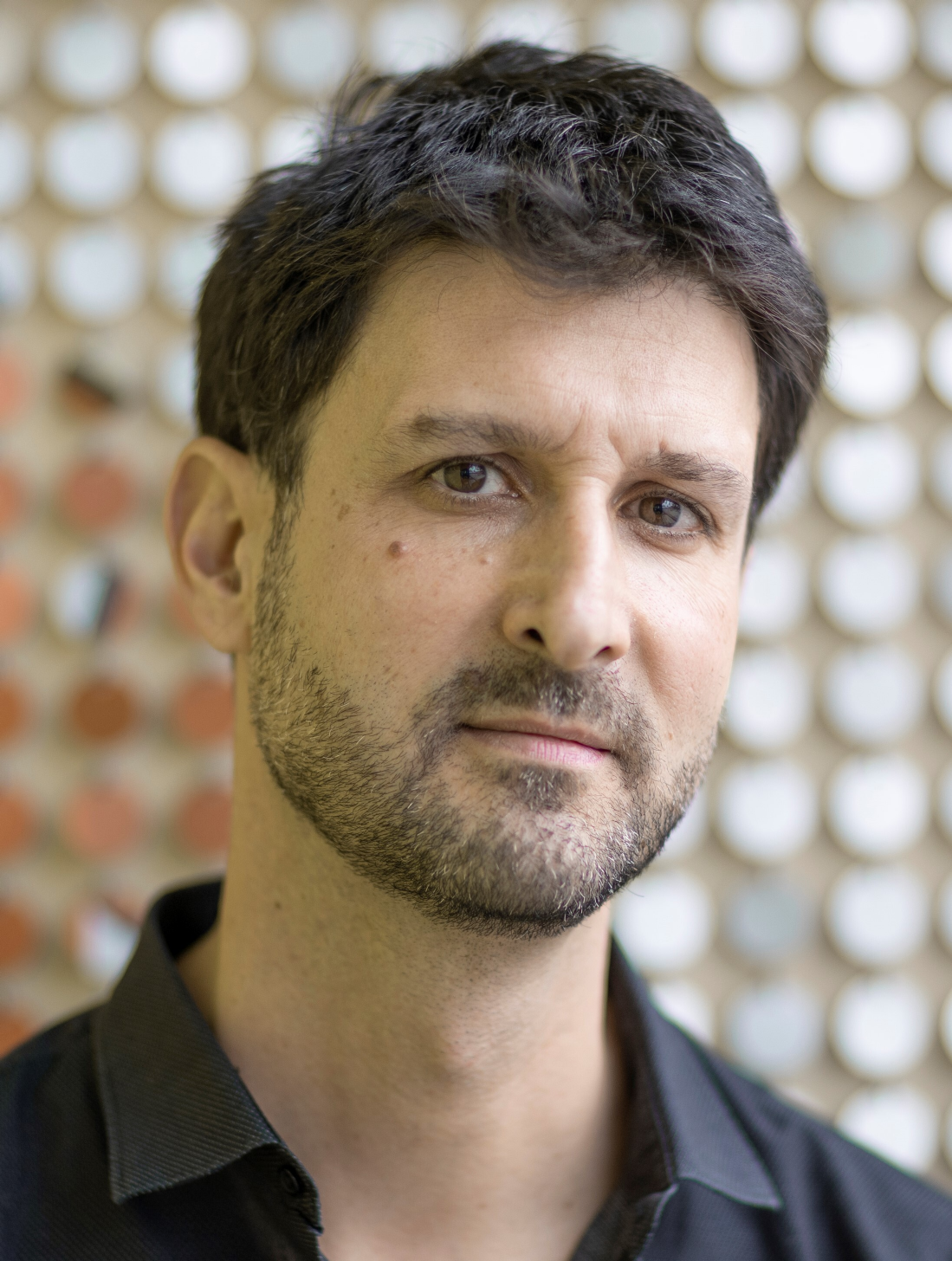}}]{Anatole Lécuyer}
is director of research and head of Hybrid team at Inria, Renne, France. He is currently Associate Editor of IEEE Transactions on Visualization and Computer Graphics, Frontiers in Virtual Reality and Presence. He was Program Chair of IEEE VR 2015-2016 and General Chair of IEEE ISMAR 2017. Anatole Lécuyer obtained the IEEE VGTC Technical Achievement Award in Virtual/Augmented Reality in 2019.
\end{IEEEbiography}

\vspace*{-3\baselineskip}







\end{document}